\documentclass[pra,twocolumn,showpacs]{revtex4}
\usepackage{amsmath, amsfonts, amssymb,bbm}
\usepackage{dsfont}                           
\usepackage[usenames]{color}                  
\usepackage{floatflt}                         
\usepackage{subfigure}
\usepackage{graphicx}                         
\usepackage{epstopdf}                         
\usepackage[colorlinks=true,
            urlcolor=blue,
            citecolor=red,
            pdfstartview=FitH,
            pdfpagemode=None]{hyperref}

\newcommand{\ket}[1]{\lvert#1\rangle}

\newcommand{\Tr}[1]{\operatorname{Tr}\bigl[#1\bigr]}
\DeclareMathOperator{\sgn}{sgn}

\begin{document}
\title{Efficient feedback controllers for continuous-time quantum error correction}

\author{Bradley A. \surname{Chase}}
\email{bchase@unm.edu}
\author{Andrew J. \surname{Landahl}}
\email{alandahl@unm.edu}
\author{JM \surname{Geremia}}
\email{jgeremia@unm.edu}
\affiliation{Center for Advanced Studies, Department of Physics and Astronomy,
             University of New Mexico,
             Albuquerque, NM, 87131 USA}

\begin{abstract}

We present an efficient approach to continuous-time quantum error correction
that extends the low-dimensional quantum filtering methodology developed by van Handel and
Mabuchi [quant-ph/0511221 (2005)] to include error recovery operations in
the form of real-time quantum feedback.  We expect this paradigm to be useful for systems in which error recovery operations cannot be applied instantaneously.  While we could not find an exact
low-dimensional filter that combined both continuous syndrome measurement
and a feedback Hamiltonian appropriate for error recovery, we developed an
approximate reduced-dimensional model to do so.  Simulations of the
five-qubit code subjected to the symmetric depolarizing channel suggests
that error correction based on our approximate filter performs essentially identically 
to correction based on an exact quantum dynamical model.  

\end{abstract}
\pacs{03.67.Pp,03.65.Yz,42.50.Lc,02.30.Yy}
\maketitle

\section{Introduction}

Quantum error correction is inherently a feedback process where the error
syndrome of encoded qubits is measured and used to apply conditional
recovery operations \cite{Gottesman:1997a}.  Most
formulations of quantum error correction treat this feedback process as a
sequence of discrete steps.  Syndrome measurements and recovery operations
are performed periodically, separated by a time-interval chosen small enough
to avoid excessive accumulation of errors but still comparable to the time
required to implement quantum logic gates
\cite{Gottesman:1997a,Nielsen:2000a}.  There is, however, mounting evidence
from the field of real-time quantum feedback control
\cite{Wiseman:1994a,Armen:2002a,Bouten:2006a,Cook:2007a} that continuous
observation processes offer new, sometimes technologically advantageous,
opportunities for quantum information processing.

Toward this end, Ahn, Doherty and Landahl (ADL) \cite{Ahn:2002a} devised a
scheme to implement general stabilizer quantum error correction
\cite{Gottesman:1997a} using continuous measurement and feedback.
Unfortunately an exact implementation of the ADL scheme is computationally
demanding.    For an $n$-qubit code, the procedure requires one to time-evolve 
a $2^n$-dimensional density matrix for the logical qubit alongside the
quantum computation \cite{Ahn:2002a}.    This classical
information-processing overhead must be performed to interpret the continuous-time
error syndrome measurement data and determine how recovery operations, in
the form of a time-dependent feedback Hamiltonian, should be applied.  
While $n$ is a constant for any particular choice of code, even modest codes such as the
five-qubit code \cite{Bennett:1996a,Laflamme:1996a} and the seven-qubit Steane code
\cite{Steane:1996a} push classical computers to their limits.  Despite
state-of-the art experimental capabilities, it would be extremely
difficult to implement the ADL bit-flip code in practice.  Consequently, Ahn and others have devised alternate feedback protocols which are less demanding \cite{Sarovar:2004a, Ahn:2004a}, but perform worse than the the original ADL scheme.

Recently, van Handel and Mabuchi addressed the computational overhead of
continuous-time error syndrome detection \cite{VanHandel:2005a} using
techniques from quantum filtering theory
\cite{Belavkin:1999a,VanHandel:2004a,Bouten:2005a}.  They developed an exact,
low-dimensional model for continuous-time error syndrome measurements, but did not
go on to treat continuous-time recovery.  The complication
is that any feedback Hamiltonian suitable for correcting errors during the
syndrome measurements violates the dynamical symmetries that
were exploited to obtain the low-dimensional filter in Ref.
\cite{VanHandel:2005a}.  While one might address this complication by simply
postponing error recovery operations until a point where the 
measurements can be stopped, there may be scenarios where it would be preferable to perform
error recovery in real-time.  For example, if the recovery operation is not instantaneous, responding to errors as they occur might outperform
protocols where there are periods without any error correction.  

In this paper, we extend the quantum filtering approach developed by van
Handel and Mabuchi to include recovery operations.  We consider an
error-correcting feedback Hamiltonian of the form devised by Ahn, Doherty
and Landahl, but our approach readily extends to other forms for the
feedback.  While an exact low-dimensional model for continuous-time
stabilizer generator measurements in the presence of feedback does not
appear to exist, we devise an approximate filter that is still
low-dimensional, yet sufficiently accurate such that high-quality error
correction is possible.  

\section{Continuous-Time Quantum Error Correction}

For our purposes, a quantum error correcting code is a triple $(E,
\mathcal{G}, R)$.  The quantum operation $E:\mathbb{C}^{2k} \mapsto
\mathbb{C}^{2n}$ \emph{encodes} $k$ logical qubits in $n$ physical qubits.
$\mathcal{G}$ is a set of $l=n-k$ stabilizer generator observables with
outcomes $\pm 1$ that define the \emph{error syndrome}.   $R:\{\pm
1\}^{\otimes l}\mapsto \mathbb{C}^{2n\times 2n}$ is the
\textit{recovery operation}, which specifies what correction should be
applied to the physical qubits in response to the syndrome measurement
outcomes. 

The particular choice of code $(E, \mathcal{G},R)$ is usually made with
consideration for the nature of the decoherence affecting the physical
qubits \cite{Knill:2000a}.  For example, the bit-flip code (considered by
both ADL and van Handel and Mabuchi) improves protection against an error
channel that applies the Pauli $\sigma_x$ operator to single qubits at a 
rate $\gamma$.  Here, we adopt the notation that  $X_n$ represents the Pauli
$\sigma_x$ operator on qubit $n$, and similarly for $Y_n$ and
$Z_n$.  In the bit-flip code,  $E$ encodes $k=1$ qubits in $n=3$ qubits by
the map $\alpha\ket{0}+\beta\ket{1} \mapsto \alpha\ket{000} +
\beta\ket{111}$.  The $l=2$ stabilizer generators are $g_1 = ZZI :=
\sigma_z\otimes\sigma_z\otimes I$ and $g_2 = IZZ :=  I \otimes \sigma_z
\otimes \sigma_z$; each extracts the parity of different qubit pairs.  The
recovery $R$, given the outcomes of measuring $(g_1,g_2)$, is defined by
$(+1,+1)\mapsto I $, $(+1,-1)\mapsto X_3$, $(-1,+1)\mapsto X_1$ and
$(-1,-1)\mapsto X_2$.  

In this paper, we focus primarily on the five-qubit-code ($n = 5, k = 1$)
that increases protection against general separable channels, and in
particular the continuous-time symmetric depolarizing channel that applies
all three Pauli operators to each of the physical qubits at the same rate
$\gamma$.  The five-qubit code has $l=4$ stabilizer generators
$\{XZZXI,IXZZX,XIXZZ,ZXIXZ\}$.  It is also a \emph{perfect} code in that all
16 distinct syndrome outcomes indicate distinct errors: one corresponding to
the no-error condition, and one syndrome for each of the three Pauli errors
on each of the five qubits.  We defer to \cite{Nielsen:2000a,
Gottesman:1997a} for the encoding and recovery procedures for this code.

\subsection{Continuous-time Stabilizer Generator Measurements}

Quantum error correction can be extended to continuous time by replacing
discrete measurements of the stabilizer generators $g_1, \ldots, g_l$ with a
set of $l$ continuous observation processes \cite{Ahn:2002a}.  This creates
$i=1,\ldots, l$ measurement records \begin{equation} dQ_t^{(i)} =
2\sqrt{\kappa} \Tr{g_i \rho_t}dt + dW_t^{(i)} \end{equation} obtained from
the encoded qubit with state $\rho_t$ at time $t$.  Here, $\kappa$ is a
constant called the measurement strength that depends upon the physical
implementation of the continuous measurement and the $dW^{(i)}_t$ are
independent Wiener processes, each with $\mathbbm{E}[dW_t] = 0$ and $dW_t^2 = dt$ \cite{Wiseman:1993a,VanHandel:2004a}.   We do not consider here
how one might implement the set of $l$ simultaneous stabilizer generator
observations other than to comment that doing so in an AMO technology would
likely involve coupling the $n$ physical qubits to a set of electromagnetic
field modes and then performing continuous photodetection on the scattered
fields.  While this model is rather general, we take the same $\kappa$ for each
qubit, implying symmetric coupling of the qubits. 

By itself, the continuous measurement record is too noisy to permit quantum
error correction---one must first process the measurement data to deal with
the presence of the noises $dW_t^{(i)}$.  The most straightforward approach
toward filtering the noise is via an estimate of the full logical qubit
density operator $\rho_t$.  Generating a full state estimation based on the evolving syndrome
measurement data is accomplished by the quantum filtering equation (in its
adjoint form with $\hbar=1$) \cite{Bouten:2005a}
\begin{eqnarray}   \label{eq:BelavkinFilter}
	d\rho_t & = & 
		\gamma  \sum_{m=1}^n \sum_{j}\mathcal{D}[\sigma_{j}^{(m)}]\rho_t dt 
		+ \kappa \sum_{i=1}^{l}\mathcal{D}[g_i]\rho_t dt \nonumber \\
	& & + \sqrt{\kappa}\sum_{i=1}^l \mathcal{H}[g_i]\rho_t \left( dQ_t^{(i)}
		 - 2 \sqrt{\kappa}\, \mathrm{Tr}[ g_i \rho_t ] dt \right) \nonumber \\
	& & 
		- i [ H_t, \rho_t ] dt\, ,
\end{eqnarray}    
where $j\in \{ x, y, z\}$ and the superoperators are defined as:
$\mathcal{D}[\sigma]\rho = \sigma\rho \sigma-\rho$ and $\mathcal{H}[g_l]\rho
= g_l\rho  + \rho g_l - 2\Tr{g_l\rho }\rho $.   The first term in the
filtering equation accounts for the action of the continuous-time symmetric
depolarizing channel while the second accounts for the effect of coupling
the logical qubit to the field used to implement the continuous
measurements.  Conditioning the quantum state $\rho_t$ on the continuous
measurement occurs via the third term, which is driven by the innovations
processes $ dQ_t^{(i)} - 2 \sqrt{\kappa}\, \mathrm{Tr}[ g_i \rho_t ] dt$.
The time evolution $\rho_t$ generated by a particular noise realization is generally called a  \emph{trajectory}.  

The final term in Eq.\ (\ref{eq:BelavkinFilter}) describes the action of the time-dependent feedback Hamiltonian used to implement error recovery.   Following Ahn, Doherty and Landahl, we choose the feedback Hamiltonian to be of the form
\begin{equation} \label{eq:Hamiltonian}
	H_t = \sum_{m=1}^{n} \sum_j \lambda_{j,t}^{(m)} \sigma_{j}^{(m)},
\end{equation}
which corresponds to applying Pauli operators $\sigma^{(m)}_j$ to each qubit
with a controllable strength $\lambda_{j,t}^{(m)}$.   The policy for determining the feedback strengths $\lambda_{j,t}^{(m)}$ at each point in time should be chosen optimally.  Ahn, Doherty, and Landahl obtained their feedback policy by defining the \emph{codespace projector} $\Pi_0$ onto the no error states (states which are $+1$ eigenvectors of all stabilizers) and then maximizing the \emph{codespace fidelity} $\Tr{\Pi_0\rho_t}$.   Assuming a maximum feedback strength  $\lambda_{\text{max}}$, the resulting feedback policy is given by setting
\begin{equation} \label{eq:FeedbackPolicy}
	\lambda_{j,t}^{(m)} = \lambda_{\text{max}} \sgn\bigl(\Tr{-i[\Pi_0,\sigma_{j}^{(m)}] \rho_t}\bigr) \, .
\end{equation}

\subsubsection{Computational Expense}

Because this is a closed-loop strategy, the feedback controller must determine each $\lambda_{j,t}^{(m)}$ from the evolving measurement in real time.    The utility of feedback in any real setting then relies greatly upon the controller's ability to integrate the filtering equation rapidly enough to maintain pace with the quantum dynamics of the qubits.  For the five-qubit code, $1024-1$ real parameters are needed to represent the density matrix.   We found that stable numerical integration \cite{Kloeden:1992a} of even a single trajectory required approximately 36 seconds on a 2.1 GHz desktop computer ($\gamma dt \approx 10^{-5}$ over a timespan $[0,0.25\gamma]$).   This is far from adequate for use in an actual feedback controller even in state-of-the-art experiments.

Moreover, Eq.\ (\ref{eq:BelavkinFilter}) is a nonlinear filter, and for such
filters it is rarely possible to evaluate even qualitative properties
analytically.  One must then average over an appreciable number of
trajectories to find the expected behavior of quantities such as the
codespace fidelity as a function of time.  For the five-qubit code, our
integrator requires approximately 10 hours to simulate 1000 trajectories.

\subsection{Reduced-Dimensional Filters}

Considering that the syndrome measurements yield information about
correlations between qubits and not information about the individual states
of the qubits, one can imagine that  propagating the full density matrix is
excessive.  Indeed, the ADL scheme only makes use of the projection of
$\rho_t$ onto the codespace, generating the same feedback policy
regardless of which state $\rho_0$ in the codespace is initially chosen.  It
is reasonable to expect that a lower dimensional model could track solely
the information extracted from the syndrome measurements.  This is exactly
the premise used by van Handel and Mabuchi to obtain a low-dimensional model
of continuous-time stabilizer generator measurements (in the absence of
feedback) \cite{VanHandel:2005a}.   They formulate the problem as a graph
whose vertices correspond to syndromes and whose edges reflect the action of
the error model.  The filtering problem is then reduced to tracking the node
probabilities, i.e., the likelihoods for the qubit to be described by each
of the various syndrome conditions.  Dynamical transitions occur between the
syndromes due to the error channel, and the filter works to discern these
transitions from the stabilizer measurement data.

\begin{figure*}
\begin{center}
	\includegraphics{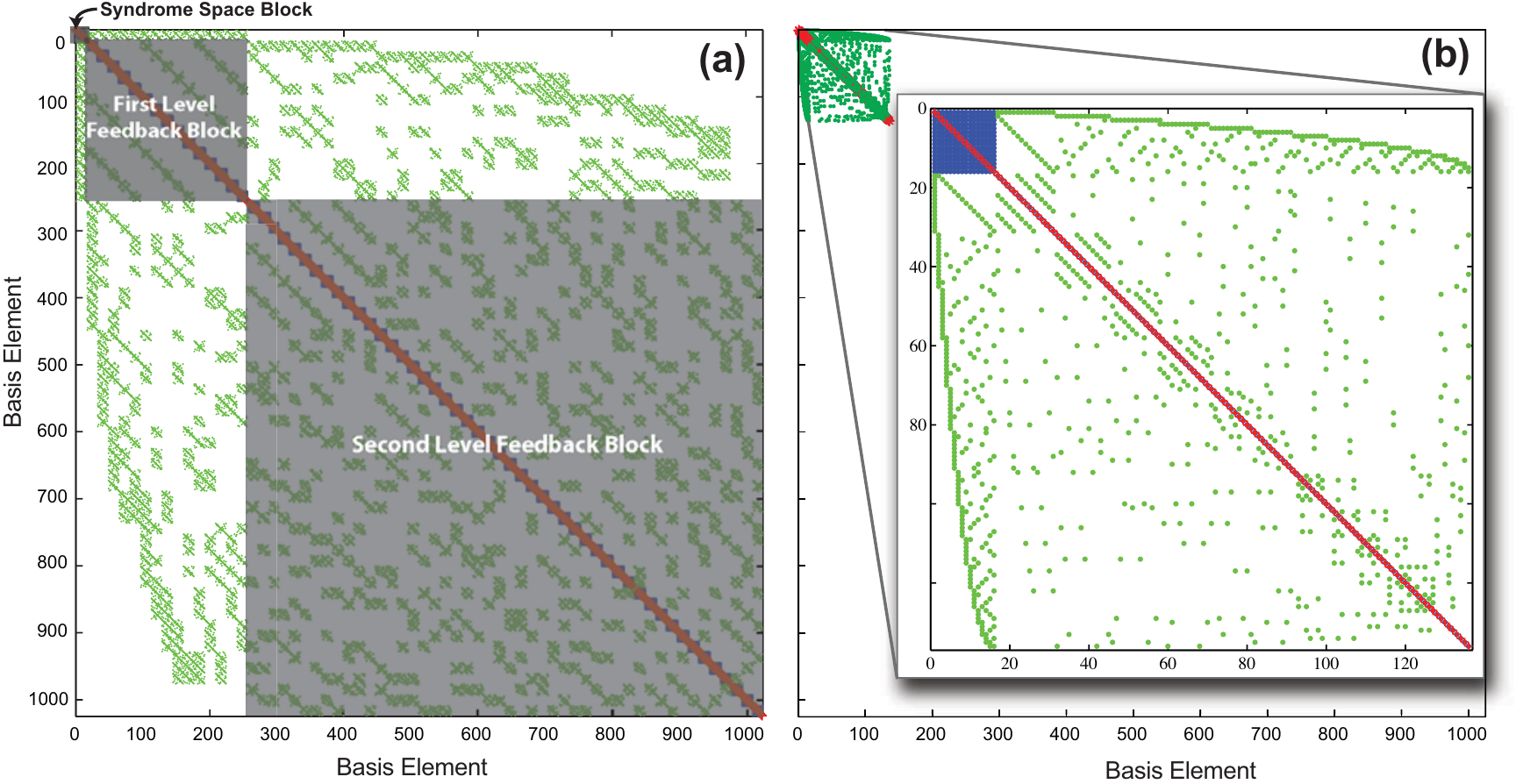}
\end{center}
\vspace{-5mm}
\caption{Non-zero matrix elements of $(a)$ untruncated and $(b)$ truncated
filter.  Blue squares correspond to decoherence terms, red crosses
correspond to measurement terms and green dots correspond to feedback terms.
Note the difference in dimension of the matrices. (Color online.)} \label{fig:annotatedMatrix}
\end{figure*}

For an $(E, \mathcal{G}, R)$ code, van Handel and Mabuchi define a set of projectors onto the distinct syndrome spaces.  For the five-qubit code, there are 16 such projectors; $\Pi_0$ is the codespace projector as before and $\Pi_{j}^{(m)}=\sigma_j^{(m)}\Pi_0\sigma_j^{(m)}$ are projectors onto states with a syndrome consistent with a $\sigma_j$ error on qubit $m$.  Forming the probabilities 
\begin{equation}
	p_{j,t}^{(m)}=\Tr{\Pi_{j}^{(m)} \rho_t}
\end{equation}
into a vector $\mathbf{p}_t$ and computing $dp_{j,t}^{(m)}$ from the full dynamics leads to the reduced filter
\begin{equation} \label{eq:BasicWonham}
	d\mathbf{p}_t = \Lambda\mathbf{p}_t \, dt 
		+ 2\sqrt{\kappa}\sum_{k=1}^{l}(H_l - \mathbf{h_l}^T\mathbf{p}_t \, I) \mathbf{p}_t \, 
			dW_t
\end{equation}
with $\Lambda_{rs} = \gamma(1-16\delta_{rs})$, $h_l^{j,m}$ the outcome of
measuring $g_l$ on $\Pi_{j}^{(m)}$ and $H_l =
\operatorname{diag}\mathbf{h}_l$ (Eq.\ (4) in Ref.\ \cite{VanHandel:2005a}).  The
equations for $p_{j,t}^{(m)}$ are closed and encapsulate all the information
that is gathered from measuring the stabilizer generators.  Equation
(\ref{eq:BasicWonham}) is an example of a \emph{Wonham filter}, which is the
classical optimal filter for a continuous-time finite-state Markov chain with an observation process driven by white noise \cite{Wonham:1965a}.    Further discussion of the Wonham filter and its use
in conjunction with discrete-time error correction can be found e.g., in Ref. \cite{VanHandel:2005a}.

\section{Continuous-Time Quantum Filtering with Feedback}

We now extend Eq.\ (\ref{eq:BasicWonham}) to include a feedback Hamiltonian suitable for error recovery.  Following van Handel and Mabuchi's lead, we see that Eq.\ (\ref{eq:BasicWonham}) was derived by taking $dp_{j,t}^{(m)} = \Tr{\Pi_{j}^{(m)} d\rho_t}$ for a basis which closed under the dynamics of the continuous syndrome measurement.  One hope is that simply adding the feedback term in by calculating $\Tr{-i\lambda_{k,t}^{(r)}\Pi_{j}^{(m)}[\sigma_k^{(r)},\rho_t]}$ also results in a set of closed equations.  However, that is not the case when using the basis of the sixteen syndrome space projectors $\Pi_{j}^{(m)}$.  Specifically,  $[\Pi_{j}^{(m)},\sigma_k^{(r)}]$ cannot be written as a linear combination of syndrome space projectors.  This is not surprising as the feedback Hamiltonian term under consideration is the only term which generates unitary dynamics.

Inspired by the form of the commutator between the feedback and the syndrome space projectors, we define feedback coefficient operators
\begin{equation} \label{eq:FeedbackCoefficient}
\Pi_{j,c}^{(m)} = (+i\text{ or } +1)\sigma^{\otimes 5}\Pi_{j}^{(m)}\sigma^{\otimes 5}\, ,
\end{equation}
where $c$ is an arbitrarily chosen index used to distinguish the $i$ or 1 prefactor and combination of Pauli matrices which sandwich the syndrome space projector $\Pi_{j}^{(m)}$.   For the five-qubit code, the syndrome projectors are simply those operators which have the 1 prefactor and 10 identity matrices.  The corresponding feedback coefficient is $p_{j,c}^{(m)}=\Tr{\Pi_{j,c}^{(m)}\rho_t}$.  If we then iterate the dynamics of the filter (\ref{eq:BelavkinFilter}) by calculating $p_{j,c}^{(m)}$ starting from the syndrome space projectors, we find that each feedback Hamiltonian term generates pairs of feedback coefficient terms.  For example, calculating the dynamics due to feedback $X_1$ on $\Pi_0$ generates two feedback coefficient operators: $\Pi_{0,0}=i\Pi_0X_1$ and $\Pi_{0,1}=iX_1\Pi_0$.  We must then determine the dynamics for these first level feedback coefficients.  This will include calculating the $Y_5$ feedback on $\Pi_{0,1}$, which generates second level feedback coefficients $\Pi_{0,2}=X_1Y_5\Pi_0$ and $\Pi_{0,3}=X_1\Pi_0Y_5$.  Continuing to iterate feedback coefficient terms, we find that an additional 1008 distinct $p_{j,c}^{(m)}$ terms are needed to close the dynamics and form a complete basis.  Adding in the initial 16 syndrome space projectors gives a 1024 dimensional basis---clearly no better than propagating the full density matrix.  However, it is now relatively easy to calculate the feedback strengths, which depend only on pairs of first-level feedback coefficients.  For example, from Eq.\ (\ref{eq:FeedbackPolicy}) we find that $\lambda_{0,t}^{(1)}=\lambda_{\text{max}}\sgn\left(-p_{0,0}+p_{0,1}\right)$, where $p_{0,0}=\Tr{\Pi_{0,0}\rho_t}$ and $p_{0,1}=\Tr{\Pi_{0,1}\rho_t}$ are first-level coefficients developed earlier in the paragraph.

\subsection{Approximate Filter for the Five-Qubit Code}
Although the dimension of the alternate basis is no smaller than the dimension of the full density matrix, the structure of the filter represented in the alternate basis provides a manner for interpreting the relative importance of the $p_{j,c}^{(m)}$ feedback coefficients.  This is best seen graphically in Fig.\ \ref{fig:annotatedMatrix}(a), which superimposes the non-zero matrix elements coming from the noise, measurement and feedback terms.  Both measurement and noise are block diagonal as expected; it is the feedback that couples blocks together in a hierarchical fashion.  This hierarchy can be parameterized by the number of ``feedback transitions'' which connect a given feedback coefficient to the syndrome space block.  For example, the upper left block, which corresponds to the syndrome space projectors, is connected via feedback terms to the first level feedback block, whose feedback coefficients are each one feedback transition away from the syndrome space block.  In turn, the first level block is then connected to a second level feedback block, whose feedback coefficients are two feedback transitions away from the syndrome block.

Given that the initial state starts within the codespace and given that
feedback is always on, the feedback coefficients that are more than one
feedback transition from the syndrome space block should be vanishingly
small.  Limiting consideration to these first two blocks, we also find that
pairs of feedback coefficients couple identically to the syndrome space
block.  For example, we find that $-iX_1\Pi_0$ and $i\Pi_0X_1$ couple to syndrome space projectors identically.  This is not surprising, as these two terms comprise the commutator that results from the $X_1$ feedback Hamiltonian.   However, outside the first level of feedback transitions, the matrix elements of these feedback coefficients differ.  Additionally, feedback coefficients involving feedback Hamiltonians which correspond to a syndrome error on the codespace projector are related as
\begin{equation} \label{eq:PauliRelation}
-i\sigma_j^{(m)}\Pi_0+i\Pi_0\sigma_j^{(m)} = -i\Pi_{j}^{(m)}\sigma_j^{(m)}+i\sigma_j^{(m)}\Pi_j^{(m)}\, .
\end{equation}
For the feedback coefficient examples just mentioned, this relation is $-iX_1\Pi_1^{(1)} + i\Pi_1^{(1)}X_1 = -i\Pi_0X_1 + iX_1\Pi_0$.  Truncating the dynamics to include only the first level of feedback and combining distinct feedback coefficients which act identically within this block results in the matrix of Fig.\ \ref{fig:annotatedMatrix}(b) over only 136 basis elements.  Note that the controller now only needs to reference a single basis element for calculating a given feedback strength $\lambda_{j,t}^{(m)}$.

\subsection{Approximate Filter for General Codes}

Our truncation scheme generalizes for reducing the dimensionality of the
quantum filter for an arbitrary $(E,\mathcal{G},R)$ code.  Such a filter for
an $[\![n,k]\!]$ quantum error-correcting code \cite{Nielsen:2000a} has the same form as
Eq.\ (\ref{eq:BelavkinFilter}), but involves $n$ physical qubits and $l = n-k$
continuous-time stabilizer generator measurements.  In the following, we assume the continuous-time symmetric depolarizing channel, though it should be straightforward to extend to other noise models.  For a non-perfect, non-degenerate code, there are a total of $2^{n-l}$ stabilizer generator measurement outcomes, but only $3n+1$ will be observed for the given noise channel.  For a perfect, non-degenerate code ($2^{n-l}=3n+1$), all possible syndrome outcomes are observed.  In either case, given the observable syndrome outcomes, we can define $3n+1$ syndrome space projectors and $3n$ feedback parameters needed for recovery.  Degenerate codes require fewer than $3n$ recovery operations, as distinct actions of the noise channel give rise to identical errors and recovery operations.  The degeneracy depends greatly on the particular code, so we merely note that degenerate codes will require \emph{fewer} syndrome space projectors and feedback parameters than their non-degenerate relatives. 

\begin{figure}[t!]
\vspace{2mm}
\begin{center} 
\includegraphics{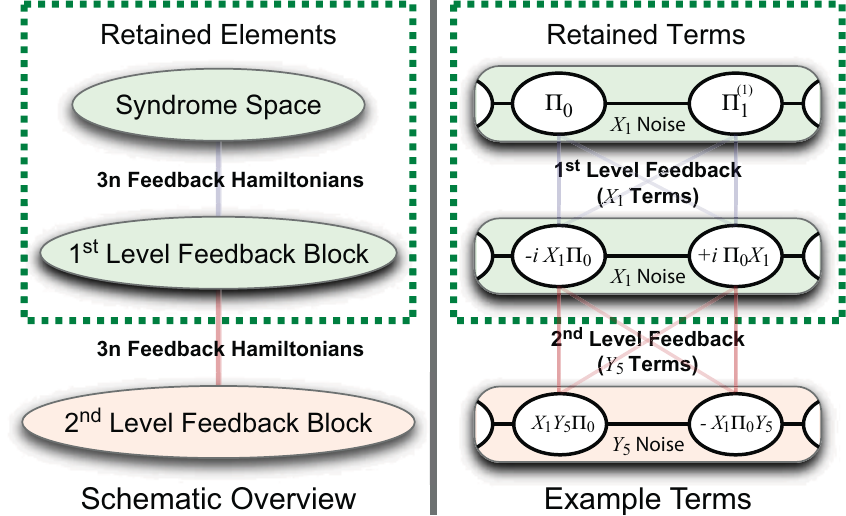}
\end{center}
\vspace{-3mm}
\caption{On the left, a schematic diagram of truncating the filter to only syndrome space and first level feedback blocks.  On the right, just a few of the 1024 feedback coefficients of the five-qubit code representing the different feedback block levels.} \label{fig:hierarchy}
\end{figure}

Once we determine the syndrome space projectors and feedback parameters for the code, we can introduce feedback coefficient operators of the
form of (\ref{eq:FeedbackCoefficient}) but over $n$ qubits.  A truncated filter is constructed as follows.
\begin{enumerate}
\item Close the dynamics of the $3n+1$ syndrome space projectors by introducing $6n(3n+1)$ first-level feedback terms (2 feedback coefficients per commutator in each of the $3n$ feedback Hamiltonians).
\item Close the dynamics of the first-level feedback terms, truncated to a basis of syndrome space and first-level feedback terms, i.e. ignore any second-level terms which were not defined in step 1.  The procedure involves $\mathcal{O}(n^3)$ steps,  since one must at least examine each of the $6n$ feedback commutator terms for all first-level feedback coefficients.
\item Each of the $3n+1$ syndrome space projectors is related to pairs of first-level feedback coefficients that comprise a commutator.  There is an additional factor of degeneracy between syndrome space projectors and feedback coefficients which involve the same Pauli matrix [c.f., Eq.\ (\ref{eq:PauliRelation})].  A similarity transform is used to eliminate these redundancies, leaving $(3n+1)+(3n+1)6n/4=\frac{1}{2}\left(2+9n(n+1)\right)$ basis elements in the fully truncated filter.
\end{enumerate}
The truncated filter requires only $\mathcal{O}(n^2)$ basis elements, as compared to the $4^n$ parameters for the full density matrix.  Additionally, the feedback strengths in Eq.\ (\ref{eq:FeedbackPolicy}) are readily calculated from the combined first-level feedback coefficients.  The truncation process is depicted schematically in the left half of Fig.\ \ref{fig:hierarchy}.  The right half of the figure gives examples of a few of the 1024 terms involved in the truncation procedure for the five-qubit code.

\begin{figure*}[t!]
\begin{center} 
\includegraphics{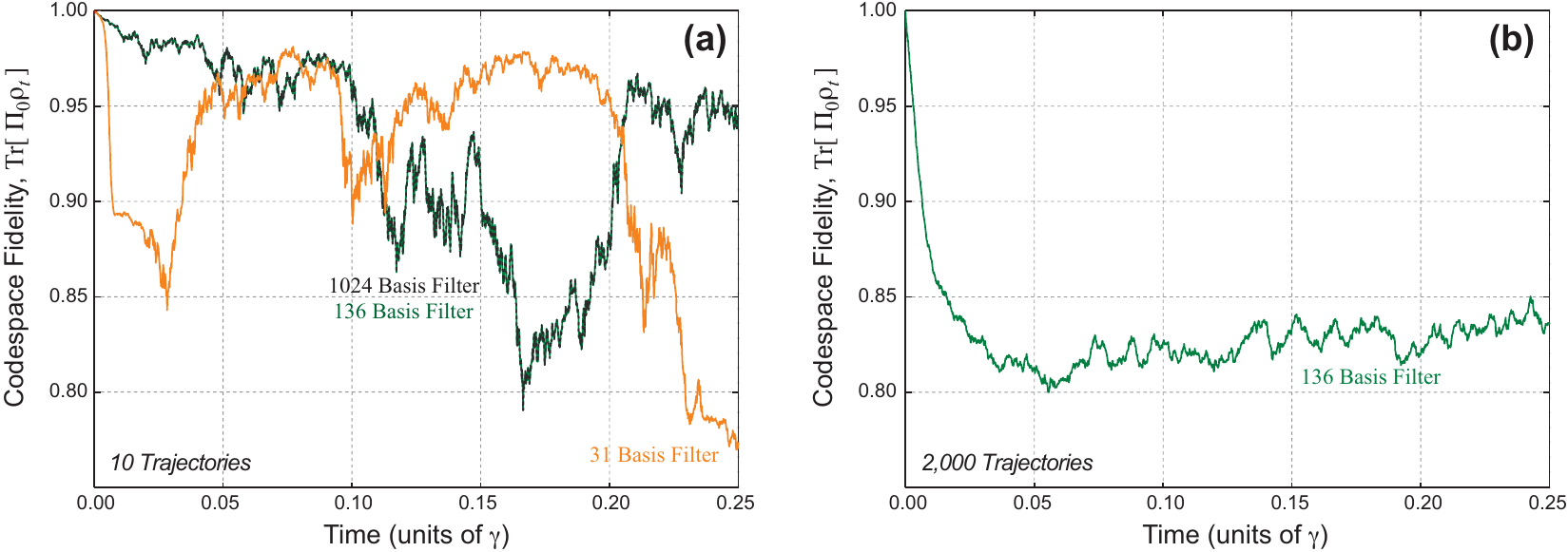}
\end{center}
\vspace{-5mm}
\caption{Numerical simulations of the five qubit code to assess the average
code space fidelity.  Plot (a) compares the codespace fidelity (averaged
over 10 trajectories) for filters with different levels of truncation: the
full (1024-dimensional) and first-level truncated (136-dimensional) filters
are essentially identical.  Plot (b) shows the codespace fidelity averaged
over 2,000 trajectories using the truncated 136-dimensional filter for error
correction.  (Simulation parameters: $\lambda_\mathrm{max} = 200 \gamma$ and
$\kappa = 100 \gamma$.)  (Color online.) \label{fig:codewordPlots}}
\end{figure*}

\subsection{Numerical Simulation}

Since the truncated filter is also nonlinear, it is difficult to provide analytic bounds on possible degradation in performance. However, we can easily compare numerical simulation between feedback controllers which use the full or truncated filter.  In fact, the dynamics should be close for the same noise realizations, indicating that they should be close per trajectory.  

In order to analyze the feedback controller's performance, the full filter Eq.\ (\ref{eq:BelavkinFilter}) is used to represent the underlying physical system.  The feedback controller was modeled by simultaneously integrating the truncated filter, driven by the measurement current from the full filter.  The feedback controller then calculated the feedback strengths which were fed back into the full filter.  The dynamics described by the full filter were then used to compute the codespace fidelity.  Using a predictor-corrector stochastic differential equation (SDE) integrator \cite{Kloeden:1992a} and varying $\kappa$ and $\lambda_{\text{max}}$ over a wide range,  we found essentially indistinguishable performance between the full and truncated filters.  Using $\kappa=100\gamma$ and $\lambda_{\text{max}}=200\gamma$ as representative parameters, Figure \ref{fig:codewordPlots}(a) demonstrates this general behavior by comparing the average codespace fidelity of a handful of trajectories using the different filters. Integrating an individual trajectory takes approximately 39.5 seconds using a 2.1 GHz PowerPC processor.  Integrating the full filter alone takes approximately 36 seconds, while integrating the  truncated filter alone takes approximately 3.5 seconds.

In addition to showing the identical performance of the full and truncated filters, Fig.\ \ref{fig:codewordPlots}(a) also shows the loss in performance if one were to truncate further.  The 31 dimensional filter is comprised of the 16 syndrome projectors and the 15 feedback coefficients which have non-zero feedback matrix elements with the codespace $\Pi_0$.  These are the only elements explicitly needed to calculate the feedback strengths in Eq. \ref{eq:FeedbackPolicy}.  This filter fails because it tacitly assumes the action of feedback on the codespace is more ``important'' than on the other 15 syndrome spaces.  Since feedback impacts all syndrome spaces equally, we need to retain those terms in order to properly maintain syndrome space probabilities. Intuitively, this suggests that the 136 dimensional filter is the best we can do using this heuristic truncation strategy.  For reference, Fig.\ \ref{fig:codewordPlots}(b) shows the average codespace fidelity of 2000 trajectories when using the truncated filter. 
\subsubsection{Comparison with Discrete Error Correction}
Given the success of the truncation scheme, we now compare the performance
of feedback-assisted error correction to that of discrete-time error correction for the five-qubit code.  The discrete model considers qubits exposed to the depolarizing channel 
\begin{equation}
d\rho_{\text{discrete}} = \gamma\sum_{j=x,y,z}\sum_{m=1}^{n=5}\mathcal{D}[\sigma_j^{(m)}]\rho_{\text{discrete}}dt
\end{equation}
up to a time $t$, after which discrete-time error correction is performed.  The
solution of this master equation can be explicitly calculated using the ansatz
\begin{equation}
\rho_{\text{discrete}}(t) = \sum_{e=0}^5\sum_{P; pw(P) = e}a_e(t)P\rho_0P ,
\end{equation}
where $P$ is a tensor product of Pauli matrices and the identity.  The
function $pw(P)$ gives the Pauli weight of a matrix, defined as the number
of $\sigma_x, \sigma_y,$ and $\sigma_z$ terms in the tensor representation.
Thus, $a_0(t)$ is the coefficient of $\rho_0$ and similarly $a_1(t)$ is the
coefficient of all single qubit errors from the initial state, e.g.,
$XIIII(\rho_0)XIIII, IIZII(\rho_0)IIZII$.  

The codespace fidelity considered earlier is not a useful metric for
comparison, as discrete-time error correction is guaranteed to restore the
state to the codespace. Following Ahn, Doherty and Landahl, we instead use the \emph{codeword
fidelity} $F_{cw}(t) := \Tr{\rho_0\rho(t)}$, which is a measure relevant for
a quantum memory.  Since error correction is independent of the encoded
state, we choose the encoded $\ket{0}$ state as a fiducial initial state.
Given that the five-qubit code protects against only single qubit errors, we
find that after error correction at time $t$, the codeword fidelity for
discrete-time error correction is
\begin{multline}
F_{cw}^{\text{discrete}} = a_0(t) + a_1(t) \\= \frac{1}{256} e^{-20 t \gamma } \left(3+e^{4 t \gamma }\right)^4 \left(-3+4 e^{4 t \gamma }\right)\, ,
\end{multline}
which asymptotes to $1/64$.  This limit arises because prior to the stabilizer generator measurements, the noise pushes the state to the maximally mixed state, which is predominately composed of the $a_2(t)$ through $a_5(t)$ terms. 

\begin{figure}[t]
\begin{center} 
\includegraphics{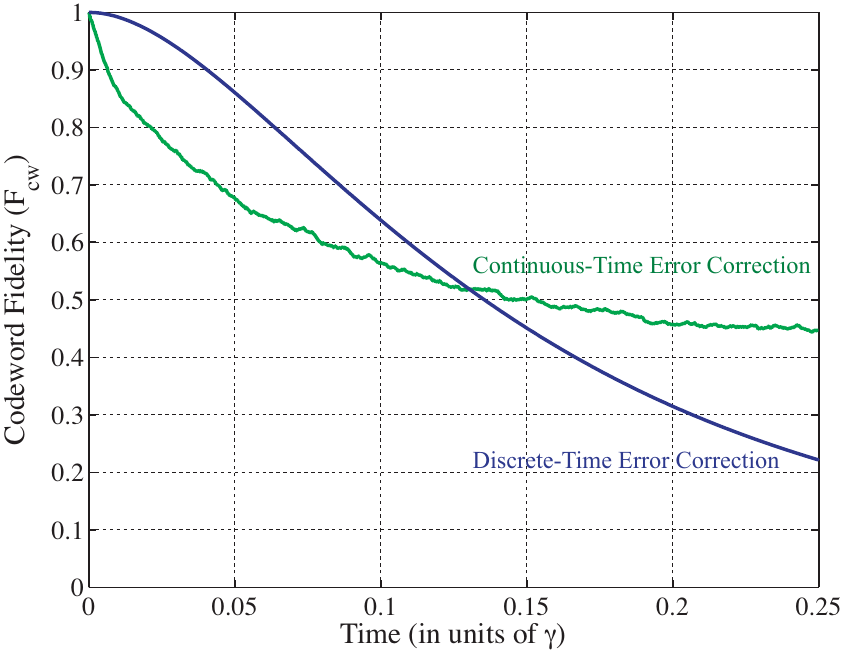}
\end{center}
\vspace{-5mm}
\caption{Comparison between continuous-time and discrete-time error correction
for the five-qubit code.  For the  continuous-time error correction
simulations, the codeword fidelity was averaged over 2,000 trajectories with
$\kappa = 100 \gamma$ and $\lambda_\text{max}=200 \gamma$. (Color online.) \label{fig:fiveQubitPerformance}}
\end{figure}

The feedback codeword fidelity $F_{cw}^{\text{feedback}}$ was calculated
by integrating both the full quantum filter (\ref{eq:BelavkinFilter}),
representing the underlying system of qubits, and the truncated filter,
representing the feedback controller.  Again, we chose $\kappa = 100\gamma$,
$\lambda_{\text{max}}=200\gamma$ and $dt = 10^{-5}\gamma$ and used the same
SDE integrator described above. Figure \ref{fig:fiveQubitPerformance} shows
the average of $F_{cw}^{\text{feedback}}$ over 2000 trajectories,
demonstrating that there are regimes where feedback-assisted error
correction can significantly outperform discrete-time error correction.  Feedback-assisted error correction appears to approach an asymptotic codeword fidelity greater than what would be obtained by decoherence followed by discrete-time error correction.  Due to the nonlinear feedback, we were unable to calculate an analytic asymptotic expression for the continuous-time strategy.
Nonetheless, the improved performance for the timespan considered suggests that better quantum memory is possible using the feedback scheme.  

\section{Conclusion}

Extending control theory techniques introduced by van Handel and Mabuchi \cite{VanHandel:2005a}, we have developed a computationally efficient feedback controller for continuous-time quantum error correction.  For our truncation scheme, the dimension of the filtering equations grows as $\mathcal{O}(n^2)$ in the number of physical qubits $n$, rather than $\mathcal{O}(4^n)$ for the original Ahn, Doherty and Landahl procedure \cite{Ahn:2002a}.  By numerical simulation of the five-qubit code, we have demonstrated the viability of such a filter for a quantum memory protecting against a depolarizing noise channel.  Moreover, in all our simulations, this performance is indistinguishable from that of the computationally more demanding filter of the ADL style.  

In systems where recovery operations are not instantaneous relative to decoherence, consideration suggests that it is desirable to perform syndrome measurement, recovery, and logic gates simultaneously.  However, it is not immediately clear how gates impact the feedback controller.  Indeed, if a Hamiltonian is in the code's normalizer, the continuous-time
feedback protocol and its performance are unchanged.  Though a universal set
of such Hamiltonians can be found, it might be desirable to find universal
gates which have physically simple interactions.  Future work involves
finding such gate sets and developing a framework for universal quantum
computing.  Additional issues of fault-tolerance and robustness could then
be explored within such a universal setup.  Exploring feedback error
correction in the context of specific physical models will provide
opportunities to tailor feedback strategies to available control parameters
and salient noise channels.  Such systems might allow the calculation of
globally optimal feedback control strategies.

\begin{acknowledgements}

We thank Ramon van Handel for helpful comments on the manuscript and Daniel Lidar for his input on how one might combine
continuous-time stabilizer generator measurements and quantum gates.  This
work was supported by the DOE NINE program under a contract with Sandia
National Laboratory (60071-699182).  AJL was also supported in part by the
NSF under grant (PHY-0555573).  JMG was also supported in part by the AFOSR
under grant (FA9550-06-1-0178).  

\end{acknowledgements}

\end{document}